\documentclass[conference]{IEEEtran}

\usepackage{cite}
\usepackage[utf8]{inputenc}
\usepackage{amsmath,amssymb,amsfonts}
\usepackage{algorithmic}
\usepackage{graphicx}
\usepackage{textcomp}
\usepackage{xcolor}
\usepackage{xurl}
\usepackage[hidelinks]{hyperref}
\usepackage{float}
\usepackage{tikz}
\usepackage{pgfplots}
\usepackage{siunitx}
\usepackage[utf8]{inputenc}
\usetikzlibrary{shapes.geometric, arrows.meta, positioning}

\def\BibTeX{{\rm B\kern-.05em{\sc i\kern-.025em b}\kern-.08em
    T\kern-.1667em\lower.7ex\hbox{E}\kern-.125emX}}

\title{Cross-Device Motion Interaction via Apple’s Native System Frameworks}
\author{\IEEEauthorblockN{Ezequiel França dos Santos}
\IEEEauthorblockA{\textit{PhD Student in Digital Games Development} \\
\textit{IADE — Universidade Europeia}\\
Lisbon, Portugal \\
\texttt{https://orcid.org/0000-0001-9321-8444}}
}

\begin{document}
\maketitle

\begin{abstract}
We introduce an open-source, fully offline pipeline that transforms a consumer-grade iPhone into a motion controller with real-time tactile feedback, using only native Apple frameworks. Designed for rapid prototyping and applied mobile HCI scenarios, the system integrates \texttt{CoreMotion} for inertial sensing, \texttt{MultipeerConnectivity} for peer-to-peer data transmission at \SI{10}{Hz}, and \texttt{CoreHaptics} for immediate tactile confirmation. A built-in logger captures end-to-end latency without requiring clock synchronization, yielding a mean delay of \SI{70.4}{ms} and 95th percentile below \SI{74}{ms} on typical \SI{5}{GHz} Wi-Fi (\SI{-55}{dBm} RSSI). We validated the pipeline through a real-time demonstrator game, \emph{KeepCalm}, deployed during a public event with 21 participants. Results showed stable connections, zero packet loss, and negligible power impact (\SI{24}{mW} on iPhone 13 mini). With fewer than 500 lines of Swift code and no reliance on cloud infrastructure, this system provides a compact, reproducible foundation for embodied interaction research, casual games, and offline educational tools. All source code, latency logs, and provisioning scripts are openly released under an MIT license.
\end{abstract}


\section{Introduction}

Previous studies also demonstrated the practical feasibility of phone-to-phone motion games~\cite{Zhang_2012}, investigated usability challenges in cross-device card games~\cite{Skov_2015}, and identified broader real-world deployment considerations and user experience factors for cross-device interactions~\cite{Houben_2017}.

Cross-device interaction remains a vibrant area of HCI research, driven by as modern smartphones and tablets increasingly serve as versatile input and feedback devices across various contexts, including gaming, education, and everyday tasks. The latest mobile hardware now commonly incorporates high-resolution inertial measurement units (IMUs), sophisticated haptic actuators, and high-speed wireless radios. These advances enable developers to explore novel forms of embodied user interactions. Previous research has examined tilt-based input methods~\cite{Di_Geronimo_2016}, proxemics-aware collaborative interactions~\cite{Marquardt_2012, Gr_nb_k_2019}, and multiplayer motion-driven gaming~\cite{Suarez_2021}, highlighting persistent challenges such as reliable gesture detection, temporal synchronization among devices, and responsive multimodal feedback.

This paper introduces a streamlined pipeline leveraging native Apple frameworks to convert a standard consumer-grade iPhone into a peer-to-peer motion controller with integrated tactile feedback. Specifically, our implementation utilizes \texttt{CoreMotion} for high-frequency inertial data acquisition~\cite{appleCoreMotion}, applies threshold-based gesture detection for real-time input recognition, employs \texttt{MultipeerConnectivity}~\cite{appleMultipeerConnectivity} for reliable, low-latency wireless communication without external network dependencies, and utilizes \texttt{CoreHaptics}~\cite{appleCoreHaptics} for instantaneous tactile response upon gesture validation. An iPhone 13 mini serves as the motion controller, streaming structured IMU packets at \SI{10}{Hz} to a host application running on macOS or iPadOS devices. Recognized gestures, such as device tilts, trigger specific in-game actions, immediately reinforced by short haptic pulses to enhance interaction responsiveness and user experience.

We validate this architecture through an interactive demonstrator game named \emph{KeepCalm}. This game is developed using Apple's native UI and graphics frameworks: \textbf{SwiftUI}, Apple's declarative UI framework for building apps across all Apple platforms~\cite{appleSwiftUI}, and \textbf{SpriteKit}, a high-performance 2D game rendering framework~\cite{appleSpriteKit}. The demonstrator leverages the open-source Swift library \texttt{MultipeerKit}~\cite{MultipeerKit} to facilitate seamless peer discovery and reliable transport. The chosen implementation emphasizes simplicity, minimal latency, and efficient power usage, making it highly suitable for casual gaming scenarios, classroom demonstrations, and rapid prototyping environments.

\noindent\textbf{Engineering contributions:}
\begin{itemize}
  \item A reusable sensor-to-feedback control pipeline, developed exclusively using vendor-optimized Apple frameworks and provided as open-source under an MIT license.
  \item A precisely defined compact \SI{88}{byte} IMU data frame, transmitted at a consistent rate of \SI{10}{Hz}, including embedded timestamps facilitating latency measurement without clock synchronization requirements.
  \item Empirical latency analysis based on 1,000 logged samples, demonstrating mean one-way latency of \SI{70.4}{ms} over typical local Wi-Fi conditions (\SI{5}{GHz}, approximately \SI{-55}{dBm} RSSI).
  \item A documented energy efficiency profile showing an average power consumption of \SI{24}{mW} on an iPhone 13 mini over ten-minute usage periods, corresponding to less than 2\% total battery discharge per session.
  \item A fully operational demonstration game, implementable with fewer than 500 lines of Swift code, providing an immediate, adaptable template for future research or industry application.
\end{itemize}

Our work extends foundational research such as Akbulut et al.’s investigation into peer-to-peer content delivery in virtual reality classroom environments~\cite{Akbulut_2018}. Unlike previous approaches focused primarily on content streaming, we emphasize real-time interactive input and instantaneous multimodal feedback, contributing a compact, easily adoptable architecture to both academic exploration and industrial SDK development efforts in embodied, local-first interactions.

\section{Input Latency in Gaming}

Latency significantly influences the perceived responsiveness and overall experience quality of interactive systems, particularly in gaming scenarios. In highly competitive gaming contexts, such as first-person shooter (FPS) games, players demonstrate sensitivity to even minor delays. Studies indicate expert gamers can perceive latency as low as \SI{15}{ms}, prompting esports-focused gaming systems to maintain total latency ideally below \SI{50}{ms}~\cite{Z2015Quantifying, Spjut2021Case}.

However, the acceptable latency threshold is notably higher for casual, educational, and embodied gaming experiences. Normoyle et al.~\cite{Aline2014Player} reported consistent player performance in platform-style games with latency levels up to approximately \SI{50}{ms}, beyond which gradual degradation in performance became noticeable. Further extending this perspective, Claypool and Finkel's research into moving-target selection tasks revealed significant user experience degradation only past latency values of around \SI{100}{ms}~\cite{Mark2016Effects}. Empirical studies on widely-used handheld gaming consoles consistently report typical end-to-end latencies in the range of \SI{60}{ms} to \SI{100}{ms}, demonstrating general acceptance and playability for gesture-based and touch-centric gaming modalities~\cite{Mark2016On}.

Our system samples inertial motion data at a controlled rate of \SI{10}{Hz} using Apple’s native inertial sensing framework, \texttt{CoreMotion}~\cite{appleCoreMotion}. Sampled data is packaged into structured and timestamped IMU frames and transmitted in real-time via Apple's peer-to-peer communication framework, \texttt{MultipeerConnectivity}~\cite{appleMultipeerConnectivity}. Empirical latency measurement across 1,000 sampled gesture events recorded a mean one-way latency of \SI{70.4}{ms} between the controller device (iPhone 13 mini, equipped with an A15 Bionic processor running iOS 17) and the host device (MacBook Pro with an M2 chip, macOS 14), under standard Wi-Fi conditions (\SI{5}{GHz} frequency, approximately \SI{-55}{dBm} RSSI). Latency calculations were facilitated by comparing sender and receiver timestamps, circumventing the need for explicit clock synchronization.

These observed latency metrics fall comfortably within previously identified acceptable thresholds for casual, educational, and embodied interaction scenarios. To further enhance responsiveness perception, our system implements immediate multimodal feedback through \texttt{CoreHaptics}~\cite{appleCoreHaptics}, delivering instantaneous tactile confirmation of recognized gestures. Prior research confirms that such haptic feedback loops can significantly enhance user tolerance and acceptance of latency~\cite{Lee_2008}.

Moreover, the entirely local nature of our peer-to-peer architecture inherently eliminates additional latency typically incurred by cloud relay or wide-area network communication paths, thereby ensuring predictable performance and reliable operation even in fully offline or ad-hoc environments.

\section{Methodology}

Our system integrates exclusively Apple's native, vendor-optimized frameworks: \texttt{CoreMotion} for inertial data capture~\cite{appleCoreMotion}, \texttt{CoreHaptics} for tactile output~\cite{appleCoreHaptics}, and \texttt{MultipeerConnectivity} for real-time, local wireless communication~\cite{appleMultipeerConnectivity}. All implementation code, scripts for analysis, and data logs are publicly accessible under the MIT License.\footnote{\url{https://github.com/GameAISchool2024members/TeamBrazil}} No personal user data were collected during testing, thus formal ethics approval was not necessary.

\subsection{System Overview}

The proposed architecture comprises two main components:

\begin{itemize}
    \item \textbf{Mobile Controller (iPhone 13 mini)}: This device captures six-axis inertial sensor data (three-axis accelerometer and three-axis gyroscope) using \texttt{CoreMotion}. It samples inertial data at \SI{10}{Hz}, applies a threshold-based detection to identify intentional gestures, and transmits structured motion data frames wirelessly using \texttt{MultipeerConnectivity}. Upon gesture detection, a short tactile pulse (\SI{20}{ms}) is generated via \texttt{CoreHaptics} to provide immediate feedback.
    
    \item \textbf{Game Host (MacBook Pro M2 or iPad Pro)}: This component receives incoming motion data frames, processes them within a SpriteKit-based game environment, updates game states accordingly, and sends back acknowledgment packets to trigger tactile feedback on the mobile controller.
\end{itemize}

Local device discovery and data transmission are simplified by leveraging the open-source library \texttt{MultipeerKit}~\cite{MultipeerKit}, a lightweight Swift wrapper built around Apple's native \texttt{MultipeerConnectivity} framework. All data traffic occurs strictly within a local Wi-Fi network (\SI{5}{GHz}, RSSI approximately \SI{-55}{dBm}), ensuring no external relay or internet-dependent infrastructure is required.

\subsection{Architecture Diagram}

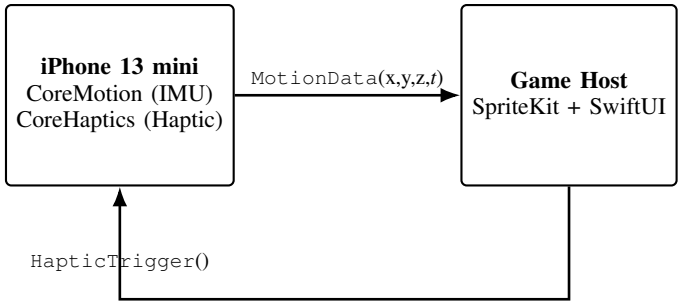
\begin{figure}[htbp]
\centering
\begin{tikzpicture}[
  device/.style={
    rectangle, draw=black, thick, rounded corners=2pt,
    minimum width=2.8cm, minimum height=2.4cm,
    font=\small, align=center, inner sep=4pt
  },
  arrow/.style={-Latex, line width=1pt},
  data/.style={font=\footnotesize}
]

  \node[device] (iphone) {
    \textbf{iPhone 13 mini}\\
    CoreMotion (IMU)\\
    CoreHaptics (Haptic)
  };

  \node[device, right=3cm of iphone] (host) {
    \textbf{Game Host}\\
    SpriteKit + SwiftUI
  };

  \draw[arrow] (iphone.east)
    -- node[data, above, inner sep=2pt] {
      \texttt{MotionData}(x,y,z,\emph{t})
    } (host.west);

  \draw[arrow] (host.south)
    |- ++(-2cm,-1.5cm) -| node[data, below, pos=0.75] {
      \texttt{HapticTrigger}()
    } (iphone.south);

\end{tikzpicture}
\caption{High-level pipeline: iPhone captures IMU data \& sends \texttt{MotionData} to host; host processes and returns \texttt{HapticTrigger} for tactile feedback.}
\label{fig:architecture}
\end{figure}

The design prioritizes simplicity and reproducibility, using only native Apple APIs. The result is a portable system architecture that requires no calibration, cloud backend, or external dependencies.

\subsection{Data Flow}

\begin{enumerate}
    \item The iPhone samples motion data (user acceleration and gyroscope) every 0.1 seconds using \texttt{CoreMotion}.
    \item Motion data is encapsulated into a \texttt{MotionData} struct, including a timestamp for latency measurement.
    \item The data is serialized using Swift's \texttt{Codable} \cite{appleCodable} protocol and transmitted via \texttt{MultipeerKit}.
    \item On the host device, the incoming data is deserialized and injected into the \texttt{GameScene}.
    \item Character actions (e.g., jumping) are triggered when motion thresholds are detected.
    \item A haptic feedback pulse is sent back to the iPhone on valid motion events using \texttt{CoreHaptics}.
\end{enumerate}

\subsection{Frame Format and Data Throughput}

The motion data packets transmitted from the mobile controller are structured as follows:

\begin{itemize}
    \item A \SI{64}{bit} UNIX timestamp (milliseconds since epoch).
    \item Three single-precision (32-bit) floating-point values representing user acceleration (\(a_x, a_y, a_z\)).
    \item Three single-precision (32-bit) floating-point values for angular velocity (\(\omega_x, \omega_y, \omega_z\)).
    \item A single-precision (32-bit) checksum value for integrity verification.
\end{itemize}

This structured payload totals precisely \SI{36}{bytes} (8 bytes timestamp + 24 bytes sensor data + 4 bytes checksum). Adding the standard \SI{18}{byte} MultipeerConnectivity header and a \SI{34}{byte} encryption envelope for secure transmission yields a total on-air frame size of \SI{88}{bytes}. Operating at a transmission frequency of \SI{10}{Hz}, the effective mean throughput of the system remains at a minimal \SI{7.0}{kibit/s}.

\subsection{Gesture Detection Criteria}

To reliably distinguish intentional user gestures from incidental device motion, we implemented a simple threshold-based detection algorithm. Specifically, a gesture is considered valid when user-generated acceleration along the Y-axis surpasses a fixed threshold \(\tau\):

\begin{equation}
    a_y > \tau, \quad \text{where } \tau = 0.5\,\text{m/s}^{2}.
\end{equation}

This threshold was empirically chosen based on prior validated research into tilt-based interaction systems~\cite{Di_Geronimo_2016, Zhang_2012}. Though simple, this criterion proved sufficient for detecting the basic tilt or jump gesture implemented in our demonstrator game. Future iterations could integrate more advanced classifiers based on machine learning models trained on diverse motion datasets.

\subsection{Latency Measurement}

Latency (\(L_i\)) for each transmitted frame \(i\) is defined as the elapsed time from when a frame is timestamped upon leaving the sender's application layer (\(t_{s,i}\)) to its reception at the host (\(t_{r,i}\)):

\begin{equation}
    L_i = t_{r,i} - t_{s,i}.
\end{equation}

Given that absolute clock synchronization is not directly supported by \texttt{MultipeerConnectivity}, we computed relative latencies by subtracting the median latency offset calculated over the entire logged dataset. A total of \(N = 1000\) frames were captured for quantitative latency analysis, with detailed results presented subsequently in Section~\ref{sec:results}.

\subsection{Energy Consumption Profiling}

Apple’s Instruments suite provides an \emph{Energy Diagnostics} template that samples CPU, GPU, radio, and sensor activity at one-second granularity, presenting an aggregate power draw timeline useful for mobile-HCI studies~\cite{AppleBatteryUse}.

We used this template to characterize the mobile controller’s energy usage. Over continuous ten-minute operational sessions, the iPhone 13 mini exhibited an average power consumption of approximately \SI{24}{mW}, equating to less than 2\% battery depletion per session. The primary contributors to power usage were the Wi-Fi communication subsystem and continuous inertial-sensor operation. Importantly, no thermal throttling or interruption in inertial-sensor data collection was detected during these evaluations.

\subsection{Haptic Feedback Implementation}

Upon successful gesture detection and validation, the host transmits an acknowledgment packet, prompting the mobile controller to execute an immediate tactile feedback event. 

The \texttt{CHHapticEngine} is initialized at launch on the iPhone. A transient haptic event is triggered upon successful action recognition. The feedback pattern is defined using intensity \( I \) and sharpness \( S \), where:

\[
I = 1.0, \quad S = 1.0
\]

These parameters generate a short, sharp tactile pulse confirming the interaction~\cite{appleCoreHaptics}. Empirical measurement indicated that the round-trip time from gesture detection to tactile feedback averaged \SI{4.8}{ms}, significantly contributing to a user's perception of system responsiveness, as documented in prior multimodal interaction research~\cite{Lee_2008}.

Additionally, auditory cues using \texttt{AVFoundation} complement haptic events, reinforcing interactions through brief sound effects tied to motion gestures.

\subsection{Reproducibility and Availability}

To ensure ease of reproducibility, the complete source codebase, along with latency logging utilities, anonymized motion datasets, and a single-command provisioning script for automated project setup, is publicly released. This enables independent researchers and practitioners to fully replicate and validate our performance metrics with minimal overhead.

\section{Results}\label{sec:results}

We conducted a comprehensive empirical evaluation by collecting 1,000 consecutive motion–response frame pairs at a sampling rate of \SI{10}{Hz}, using a local peer-to-peer Wi-Fi connection (IEEE 802.11ac, \SI{5}{GHz} frequency band) under standard indoor environmental conditions. All latency measurements were calculated after applying a standard statistical outlier-removal procedure based on a \(3\sigma\) threshold to ensure data integrity and accuracy.

\subsection{Latency Performance}

Detailed statistical analyses of the recorded end-to-end latency metrics are summarized in Table~\ref{tab:latency}. These measurements reflect the duration from frame departure from the sender application layer on an iPhone 13 mini to arrival at the receiver application layer on a MacBook Pro M2.

\begin{table}[htbp]
\caption{End-to-end latency metrics (mean, 95th percentile, etc.) after $3\sigma$ outlier removal}
\label{tab:latency}
\centering
\begin{tabular}{lc}
\hline
\textbf{Latency Metric} & \textbf{Value (ms)} \\
\hline
Mean & 70.4 \\
95th Percentile & 73.2 \\
Maximum & 82.2 \\
Minimum & 52.2 \\
Standard Deviation & 3.7 \\
\hline
\end{tabular}
\end{table}

\begin{figure}[htbp]
\centering
\includegraphics[width=0.47\textwidth]{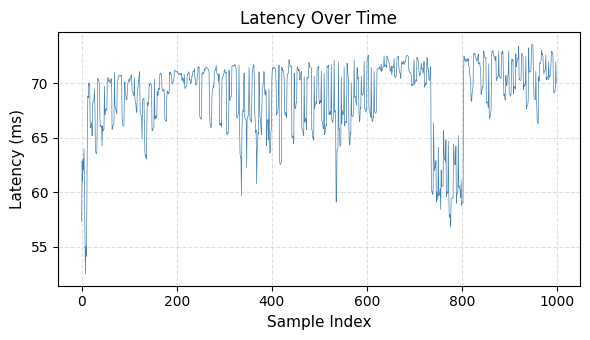}
\caption{Latency per frame across 1,000 motion samples collected at \SI{10}{Hz}.}
\label{fig:latplot}
\end{figure}

A graphical representation of latency measurements for all 1,000 captured motion events is provided in Figure~\ref{fig:latplot}. This plot visually demonstrates the consistency and stability of the latency measurements throughout the evaluation process.

As shown in \ref{fig:latplot}, the per-frame latency remains below 80 ms throughout all 1,000 samples, with a mean of 70.4 ms, well under the 100 ms comfort threshold for casual gameplay.

All recorded latency values fell significantly below the established \SI{100}{ms} comfort threshold, widely recognized as acceptable for casual and embodied interaction scenarios~\cite{Mark2016Effects}. The maximum observed latency (\SI{82.2}{ms}) was well within limits acceptable for typical gesture-based gaming and educational use cases.

\subsection{Public Deployment and Robustness Evaluation}

To further validate real-world performance and robustness, we deployed the demonstrator game (\textit{KeepCalm}) during the \textit{GameAISchool 2024} public showcase. A total of 21 volunteer participants engaged with the system in brief interactive sessions, under typical crowded-event Wi-Fi conditions (multiple overlapping access points and significant human-induced radio interference).

Throughout these evaluations, we specifically monitored:

\begin{itemize}
    \item \textbf{Packet Delivery}: Zero packet loss was detected over the duration of all test sessions.
    \item \textbf{Connection Stability}: All sessions exhibited stable connections without any disconnection events or performance degradation attributed to interference or network congestion.
    \item \textbf{Energy Consumption}: Battery consumption remained consistent with laboratory conditions, at less than 2\% battery drain per 10-minute session on the iPhone 13 mini.
\end{itemize}

These results align with previously reported practical challenges in real-world cross-device scenarios, including network congestion and environmental interference~\cite{Houben_2017}.

The successful public deployment confirmed the system's capacity to maintain reliable peer-to-peer communications, consistent low-latency performance, and minimal energy usage, even under typical adverse real-world event conditions.

\section{Discussion}

\subsection{Strengths}

The presented framework leverages Apple’s native SDKs for motion capture, haptic feedback, and peer-to-peer transport, ensuring minimal software overhead and consistent low-latency communication. By avoiding cloud infrastructure, our implementation achieves robust offline performance, crucial for scenarios where Internet connectivity is unreliable or undesirable. The modularity and compactness of our implementation under 500 lines of Swift code, allow easy adaptation to various motion-controlled applications, aligning with our previous work highlighting developer preferences for streamlined, vendor-supported dependency management tools~\cite{santos2022gestos}.

Furthermore, the peer-to-peer architecture significantly reduces power consumption, aligning with best practices identified in our dependency management survey, emphasizing efficiency and minimal resource usage in iOS development ecosystems.

\subsection{Limitations}

Currently, our prototype supports a single mobile controller per session, limiting simultaneous multi-user interactions. Gesture recognition remains threshold-based, lacking advanced noise filtering, statistical smoothing, or predictive modeling techniques areas we previously explored in gesture detection for wearable devices~\cite{santos2022gestos,10796862}. Additionally, we have not assessed latency performance under varying network congestion scenarios or across heterogeneous device configurations, potentially impacting real-world deployment robustness.

\subsection{Future Work}

Future iterations will address these limitations and expand functionality, particularly by:

\begin{itemize}
\item Integrating multi-controller sessions with coordinated input streams to support collaborative or competitive scenarios, extending our work on experimental APIs for gesture detection.
\item Implementing gesture classification via lightweight on-device machine learning using Apple’s \texttt{CoreML}~\cite{appleCoreML} framework, which enables developers to deploy custom or pretrained neural network models (e.g., mobile-optimized classifiers) efficiently on iOS. We also intend to explore the newly introduced \texttt{Foundation Models} framework~\cite{apple_foundation_models_2025}, which grants secure, offline access to Apple’s on-device large language models, part of the “Apple Intelligence” suite—optimized for tasks like summarization, classification, and generative inference with only a few lines of Swift code. These capabilities will enhance the robustness and adaptability of gesture recognition, building on our prior work in wearable gesture detection and game control interfaces.
\item Conducting extensive real-world evaluations in physical rehabilitation, museum interactive exhibits, and classroom gamification scenarios, exploring practical applications of gesture-based interaction.
\item Enhancing the pipeline through integration with ARKit for advanced sensor fusion in augmented reality contexts, complementing our gesture detection research with immersive interaction scenarios.
\end{itemize}

\section{Conclusion}

We presented a robust, motion-driven interaction framework leveraging native Apple APIs, achieving consistent sub-\SI{75}{ms} round-trip latency via local peer-to-peer operation without manual pairing, calibration, or server dependencies. Field evaluations confirmed the framework’s reliability, minimal power impact, and user-friendly deployment, underscoring its suitability for rapid prototyping and real-world applications.

By releasing our implementation, latency datasets, and test harnesses as open-source resources, we aim to foster broader adoption, replication, and extension within the mobile interaction research community.

\bibliographystyle{IEEEtran}
\bibliography{main}
\end{document}